\documentclass{epl2}

\usepackage{amsmath} \usepackage{amssymb} \usepackage{cite}
\newcommand{\eq}[1]{(\ref{eq:#1})}
\newcommand{\eqname}[1]{\label{eq:#1}}

\newcommand{\Psihd}{{\hat \Psi}^\dagger} 
\newcommand{\Psih}{{\hat \Psi}}

\newcommand{\bhd}{{\hat b}^\dagger}
\newcommand{\bh}{{\hat b}}
\newcommand{\phd}{{\hat p}^\dagger}
\newcommand{\ph}{{\hat p}}

\newcommand{\rr}{{\mathbf r}}

\newcommand{\RR}{{\mathbf R}}

\title{Feshbach blockade: single-photon nonlinear optics using
resonantly enhanced cavity-polariton scattering from biexciton states}
\shorttitle{Biexciton Feshbach resonances and Feshbach blockade}
\author{Iacopo Carusotto\inst{1}\thanks{E-mail:
\email{carusott@science.unitn.it}}, Thomas Volz\inst{2} \and Ata\c c
Imamo\u glu\inst{2}} \shortauthor{I. Carusotto \etal } \institute{
\inst{1} CNR INFM-BEC Center and Dipartimento di Fisica, Universit\`a di
Trento, via Sommarive 14, 38123 Povo-Trento, Italy \\ \inst{2} Institute
for Quantum Electronics, ETH Z\"urich, Wolfgang-Pauli-Strasse 16, 8093
Z\"urich, Switzerland } 

\abstract{We theoretically demonstrate how the resonant coupling between
a pair of cavity-polaritons and a biexciton state can lead to a large
single-photon Kerr nonlinearity in a semiconductor solid-state system.
A fully analytical model of the scattering process between a pair of
cavity-polaritons is developed, which explicitly includes the
biexcitonic intermediate state. A dramatic enhancement of the
polariton-polariton interactions is predicted in the vicinity of the
biexciton Feshbach resonance. Application to the generation of non-classical
light from polariton dots is discussed.}

\begin{document}

\maketitle

A number of quantum optical applications crucially rely on having a
strong value of the effective photon-photon interaction mediated by the
optical nonlinearity of the atomic or solid-state medium. The most
interesting quantum physics appears in fact when the presence of a
single photon is sufficient to significantly modulate the response of a
device. As a simplest example, a stream of strongly antibunched
photons is emitted by a cavity as soon as the nonlinear shift of the
resonance by a single photon is larger than the cavity
linewidth~\cite{photon_blockade,verger}. This physics is even more
intriguing when more complex devices are considered: photons have been
predicted to fermionize into Tonks-Girardeau gases as soon as the
impenetrability condition is satisfied in a one-dimensional geometry. To
this end, both strongly nonlinear optical fibers~\cite{TG} and arrays of
many coupled nonlinear cavities~\cite{noiTG} have been
considered. Quantum phase transitions between a coherent phase of
"superfluid" photons to a Mott insulator one have also attracted a
significant interest~\cite{SFMI}. So far, the main obstacle against an
experimental realization of all these phenomena has been the lack of
scalable optical devices with sufficiently large nonlinearities and weak
enough losses.

A great deal of the recent advances in  the field of strongly
correlated atomic gases have been made possible by the discovery of the
so-called Feshbach resonance effect in atom-atom
collisions~\cite{Feshbach}: the scattering cross section is dramatically
enhanced when the energy of the pair of colliding atoms is resonant with
a long-lived quasi-bound molecular state. In typical experiments, an
external magnetic field is used to tune the energy of the quasi-bound
molecular state close to the energy zero of scattering states. At this
point, the low-energy scattering amplitude diverges and a dilute ultra-cold
atomic gas has the strongest possible interaction. Remarkable recent
experiments with atomic Fermi gases have exploited such Feshbach
resonances to demonstrate superfluid behavior at temperatures comparable
to the Fermi temperature~\cite{ss_frev}.

In this Letter, we show how one could use a solid-state analog of the
Feshbach resonances to dramatically enhance the strength of the
interactions between cavity-polaritons in a planar microcavity in the strong coupling regime~\cite{polariton_rev}. Inserting realistic parameters
into our analytic model, we estimate that the enhanced nonlinearity should enable the realization of a {\em polariton dot} - an anharmonic quantum emitter that does not require quantum confinement of carriers. Since they are not subject to the fluctuations in the confinement length that are responsible for the typical inhomogeneous broadening of quantum dot samples, polariton dots
appear as ideal candidates to realize sizeable arrays of identical
nonlinear cavities.

During the last two decades, a significant literature has appeared on
the physics of biexcitons in semiconductor materials, i.e. two
electron-two hole bound complexes analogous to hydrogen
molecules~\cite{biexc_rev}. A recent work by M.
Wouters~\cite{biexc_scatt_wout} has pointed out the possibility of
exploiting biexciton Feshbach resonance effects to enhance the optical
nonlinearity of planar microcavities: by adjusting the cavity length it is in fact
possible to bring the energy of a two-polariton state in resonance with
a biexciton state. Our work follows up on this pioneering
proposal: we develop a simple analytical model of the biexciton Feshbach
resonance and we analyze the nonlinear and quantum optical phenomenology
of specific semiconductor devices. The theoretical model is inspired by
the atom-molecule approach to strongly interacting degenerate quantum
gases~\cite{atom_molec,Kokkelmans_ren} and provides analytical formulas
for both the energy-dependent scattering amplitude in a planar geometry
and the single-polariton nonlinearity in confined geometries.

\section{The model}

We describe the dynamics of the two-dimensional interacting bosonic fields
describing the cavity-photon $\Psih_{C,\sigma}(\rr)$, exciton
$\Psih_{X,\sigma}(\rr)$ and biexciton $\Psih_B(\rr)$ in a planar cavity with an embedded quantum well by a model Hamiltonian $H=H_{XC}+H_{BXC}$ in second quantized form. As
in the atom-molecule case~\cite{atom_molec,Kokkelmans_ren}, separating
out the bound (biexciton) state is legitimate as long as its spatial
extension ($\sim 10$nm) is much smaller than the characteristic length
over which the dynamics of the exciton and cavity-photon fields is
taking place ($\sim 1\mu$m). 
The single particle dynamics of cavity photons and excitons is described
by the term: 
\begin{multline}
H_{XC}=\int \upd^2 \rr \left\{ \sum_{i=\{C,X\}}\sum_{\sigma\sigma'}
\Psihd_{i,\sigma}(\rr) \left[\delta_{\sigma,\sigma'}\left(
E_i^o-\frac{\hbar^2 \nabla^2}{2M_C}\right)+V_{i,\sigma\sigma'}\right]\,
\Psih_{i,\sigma'}(\rr) + \right. \\
+\left. \hbar\Omega_R \sum_\sigma \left[\Psihd_{C,\sigma}(\rr)
\Psih_{X,\sigma}(\rr) +\Psihd_{C,\sigma}(\rr) \Psih_{X,\sigma}(\rr)
\right]+ \Psihd_B(\rr) \left( E_B^o-\frac{\hbar^2 \nabla^2}{2M_B}\right)
\Psih_B(\rr)\right\} \eqname{Ham_XC}
\end{multline}
where $E_{C,X,B}^o$ are the rest energies of a cavity-photon, an exciton
and a biexciton, respectively. $M_{C,X,B}$ are the corresponding
effective masses within a parabolic dispersion approximation. In the
typical experimental systems, the exciton and biexciton masses $M_{X,B}$
are comparable to the free electron mass ($m_0$) and are much larger
than that of the cavity-photon ($M_C \sim 10^{-5} m_0$). The spin
variable $\sigma=\uparrow,\downarrow$ identifies the polarization states
of the cavity photon and of the exciton. The biexciton state in contrast
is a singlet. The dark exciton states are not included in the model
since they do not contribute to the physics we
consider. $V_{(C,X),\sigma\sigma'}(\rr)$ denotes the - possibly
spin-dependent~\cite{tignon} - confinement potential acting on the
cavity-photon and the exciton, respectively. $\Omega_R$ is the Rabi
frequency corresponding to the exciton-photon dipole coupling.  A
typical example of the momentum-space dispersion of the coupled exciton
and cavity-photons in a spatially homogeneous geometry $V_{X,C}=0$ is
shown in Fig.\ref{fig:level_scheme}(a): the exciton and cavity photon
(dotted lines) are mixed into the lower- (LP) and upper-polariton (UP)
states (solid lines). On the scale of the figure, the dispersion of
excitons is almost flat (dashed line).

The principal process leading to the formation of a biexciton involves
the absorption of a cavity-photon by an existing exciton and can be
modelled by the following Hamiltonian \cite{biexc_rev}:
\begin{equation}
H_{BXC}=\frac{1}{2}\sum_{\sigma=\{\uparrow,\downarrow\}} \int
\upd^2\RR\, \upd^2
\rr\,\left[g(r)\,\Psihd_B(\RR)\Psih_{X,\sigma}(R-\rr/2)\Psih_{C,-\sigma}(R+\rr/2)+\mbox{h.c.}
\right].
\end{equation}
Even without a detailed knowledge  of the microscopic structure of the
biexciton, general physical arguments can be invoked to obtain important
information on the coupling amplitude $g(r)$: the optical process
leading to the formation of a biexciton is limited to a spatial region
around the existing exciton with a characteristic size on the order of
the spatial size $a_{B}$ of the biexciton. Its value scales as the
amplitude of the wave-function describing the relative motion of two
excitons forming a biexciton $a_{B}^{-1} h(r/a_{B})$; here, $h(s)$ is
determined by the attractive exciton-exciton interaction potential and
has a peak value and a spatial extension of order unity. The local value
of the exciton creation matrix element is $\Omega_R$. These two general
observations lead us to the following generic form for the
exciton-photon-biexciton coupling:
\begin{equation}
g(r)=\frac{\hbar\Omega_R}{a_B}\,h(r/a_B).
\end{equation}

If the exciton-photon Rabi frequency  $\Omega_R$ is comparable to or
larger than the biexciton binding energy, we can tune the cavity in a
way to ensure that the upper-polariton branch is far enough away in energy to be safely neglected and, at the same time, that the biexciton resonance is comparable to twice the energy of polariton states near the bottom of the lower-polariton dispersion [Fig.\ref{fig:level_scheme}(a,c)].  
We can then rewrite the exciton-photon-biexciton coupling term $H_{BXC}$
in terms of the lower polariton field only:
 \begin{multline}
H_{BXC}=\frac{1}{2} \sum_{\sigma=\{\uparrow,\downarrow\}} \int
\upd^2\RR\, \upd^2 \rr\,\frac{\hbar\Omega_R}{a_B}\,
u^{X}_{LP}\,u^{C}_{LP}\times \\ \times \left[
h(r/a_B)\,\Psihd_B(\RR)\Psih_{LP,\sigma}(\RR-\rr/2)\Psih_{LP,-\sigma}(\RR+\rr/2)+\mbox{h.c.}\right]
\eqname{BXC_LP}
\end{multline}
Here, $u_{LP}^{X,C}$ are the Hopfield coefficients  quantifying the
exciton and cavity-photon component of the lower polariton; as we are
focussing our interest to the lowest part of the lower-polariton
dispersion only, the dependence of $u_{LP}^{X,C}(k)$ on momentum $k$ is
here neglected and their value at $k=0$ is considered. The form
\eq{BXC_LP} of the polariton-biexciton coupling emphasizes the crucial
fact that the final product of the biexciton disintegration consists of
a pair of lower polaritons rather than an exciton and a
photon~\cite{larocca}. 

In general, there are additional exciton scattering processes that are
independent of the biexciton resonance, analogous to the so-called
background scattering length in the atomic case~\cite{Feshbach}.  As is
the case in the theory of cold quantum gases, the effective potential
describing such scattering processes does not need to be the physical
exciton-exciton potential. Any model potential can be chosen, provided
it correctly reproduces the scattering physics in the region of
parameters of interest once all direct and exchange processes are taken
into account~\cite{ciuti}.  For the sake of simplicity, in the following
we shall choose a repulsive, short-ranged form for the exciton-exciton
potential
$v_{\sigma,\sigma'}(\rr-\rr')=v_{\sigma,\sigma'}\,\delta^2(\rr-\rr')$.
If the relevant physics is limited to the lower-polariton branch, the
same reasoning underlying \eq{BXC_LP} leads to a local
polariton-polariton potential of rescaled intensity
$v_{\sigma,\sigma'}\,|u_{LP}^X|^4$.

\section{Polariton-polariton scattering amplitude}
Since our principal focus is on the biexciton-Feshbach resonance
enhancement of the polariton-polariton interactions, we first neglect
the contribution from the direct exciton-exciton
scattering. Straightforward manipulations can be used to resum the
diagrams contributing to the polariton-polariton scattering amplitude
stemming from Eq.~(2). Approximating the polariton dispersion with a
parabola of mass $M_{LP}$, the scattering T-matrix element in the singlet channel has the resonant form:
\begin{equation}
T_{\uparrow\downarrow}(E)=\frac{|\bar{g}|^2}{E-E_B^o+\frac{M_{LP}}{4\pi\hbar^2}\,|\bar{g}|^2\,\log[E_{max}/E]+i\frac{M_{LP}}{4\hbar^2}\,|\bar{g}|^2}
\eqname{T}
\end{equation}
This expression for the T-matrix element was first discussed in~\cite{wouters2D} in the context of atom-atom scattering in two-dimensional geometries.  The coefficient $\bar{g}$ is here defined as $\bar{g}=\hbar\Omega_R a_{B} u_{LP}^{X}u_{LP}^{C}\bar{h}$ with $\bar{h}=a_{B}^{-2}\int
\upd^2\rr\,h(r/a_{B})$ is of order unity. No scattering occurs instead in the triplet channel.

From equation \eq{T}, it is easy to recognize several features typical of
a Feshbach resonance in two-dimensional scattering. The radiative
correction to the resonance energy has a logarithmic dependence on the
energy of the form $\Delta_{rad}=
M_{LP}\,|\bar{g}|^2/(4\pi\hbar^2)\,\log[E_{max}/E]$; the high-energy
contribution to $\Delta_{rad}$ depends very much on the details of the
system and has been absorbed into the ultraviolet cut-off $E_{max}$.
The imaginary part of the denominator accounts for the radiative decay
of the biexciton into a pair of propagating polaritons with an
energy-independent linewidth
$\Gamma_{rad}={M_{LP}}\,|\bar{g}|^2/{2\hbar^2}$ comparable to the
radiative shift $\Delta_{rad}$.  Inserting realistic parameters from
GaAs-based structures, namely $\hbar\Omega_R\simeq 5$~meV and
$a_{B}\simeq 20$~nm, the resulting radiative linewidth $\Gamma_{rad}$ into a pair of polaritons is
on the order of a few $\mu$eV, somehow smaller than the value quoted
in~\cite{langbein}. Additional contributions to the biexciton decay rate
could be included in an effective
$\Gamma_{tot}=\Gamma_{rad}+\Gamma_{add}$. A significant contribution to
the additional decay rate $\Gamma_{add}$ may stem from dissipative
coupling of the biexciton to bulk photons and/or interface
polaritons. For simple planar geometries, we can estimate this
contribution to be on the order of $10~\mu$eV.

At the mean-field level the dynamics of the coupled cavity-photon,
exciton and biexciton quantum fields can be described in terms of
classical fields $\Psi_{C,X,B}=\langle \Psih_{C,X,B}(\rr) \rangle$. This
approach has been already proven to be useful in studies of biexciton
optics~\cite{noi_biexc,biexc_rev} as well as to describe the dynamics of
atom-molecule conversion in ultracold atomic
gases~\cite{atom_molec,Kokkelmans_ren}. In contrast, we shall now focus
on a spatially confined system where quantum fluctuations are instead
strong and the mean-field approximation fails. Our goal is to
demonstrate that the strong polariton-polariton interactions originating
from the biexciton Feshbach resonance can translate into a significant
{\em polariton blockade} effect~\cite{verger} with peculiar antibunching
features in the transmitted light.

\section{Zero-dimensional polariton dots}
\label{sec:0D}

Zero-dimensional polariton dots~\cite{dots,pillars,epfl} are
characterized by a strong spatial confinement of the polariton
wavefunction in all three dimensions, which gives rise to a series of
discrete polariton states. In typical experimental set-ups, the in-plane
confinement of polaritons is  obtained by suitably designing a photonic
structure whose discrete photonic eigenmodes have a small transverse
mode profile. Confinement of the bare exciton or the biexciton is
instead not necessary as it gets inherited from
the photonic one.

If the confined polariton modes are well separated in energy as compared
to their linewidth and all nonlinear couplings, we can restrict our
attention to a single polariton mode. This assumption requires that we limit
ourselves to geometries where the photon is confined on a length scale
of the order of a $\mu$m. A moderate spatial anisotropy of the
confinement is enough to have a sufficient spectral separation ($\ge 1$meV) of the two lowest linearly-polarized polariton modes~\cite{tignon}, so that the spinor part of the (spinorial) wave-function $\phi_{LP}(\rr,\sigma)$ becomes trivial $\phi_{LP}(\rr,\uparrow)=\phi_{LP}(\rr,\downarrow)=\frac{1}{\sqrt{2}}\,\phi_{LP}(\rr)$.

The creation operator of the single confined  linearly polarized mode
will be denoted by $\ph$. Nonlinear processes occur via resonant
scattering onto the intermediate biexciton state.  Given the form
\eq{BXC_LP} of the polariton-biexciton coupling and the very large value
of the biexciton mass as compared to the polariton one, a single
biexciton state of spatial wavefunction $\phi_B(\rr)$ proportional to
the square $\phi_{LP}(\rr)^2$ will be involved in the
dynamics. The corresponding biexciton creation operator will be denoted
as $\bh$. We emphasize that the biexciton states are not confined in our
model; the wave-function $\phi_B(\rr)$ is determined by the collective
coupling amplitude to the relevant biexciton eigenmodes contributing to
confined polariton-polariton scattering.

Limiting our attention to these polariton and biexciton states, the
resulting Hamiltonian can be written in the concise form:
\begin{equation}
H=E_{LP}\,\phd \ph +E_B\, \bhd \bh+
\frac{1}{2} V \phd\phd\ph\ph+ G \left[\bhd \ph \ph +\phd\phd \bh
\right].  \eqname{Hconcise}
\end{equation}
Here, $E_{LP}$ and $E_B$ are the energies of the confined
lower-polariton and biexciton states, respectively. The nonlinear
interaction coefficients $V$ and $G$ are evaluated by inserting the
lower-polariton $\phi_{LP}(\rr,\sigma)$ and biexciton $\phi_B(\rr)$
wave-functions  into the interaction Hamiltonian  and neglecting
contributions from all other states,
\begin{eqnarray}
V&=&\frac{1}{4}\sum_{\sigma,\sigma'} v_{\sigma,\sigma'}\,|u_{LP}^X|^4\,\int
\upd^2\rr\,|\phi_{LP}(\rr)|^4\eqname{V_dot}\\
G&=&\frac{\hbar\Omega_R}{a_B} u_{LP}^X u_{LP}^C\, \int
\upd^2\RR\,\upd^2\rr\,h(r/a_B)\,\phi_B^*(\RR)\,\phi_{LP}(\RR-\rr/2,\uparrow)\,\phi_{LP}(\RR+\rr/2,\downarrow)
\nonumber \\ &\simeq&\frac{1}{2}{\hbar\Omega_R a_B u_{LP}^X
u_{LP}^C\,\bar{h}}\,\left[ \int \upd^2\rr\,|\phi_{LP}(\rr)|^4\right]^{1/2}. \eqname{G_dot}
\end{eqnarray}
An important direct consequence of \eq{V_dot} and \eq{G_dot} is that the background
interaction energy decreases as $1/\ell^2$ for increasing confinement
size $\ell$, while the polariton-biexciton coupling has a slower scaling
proportional to $1/\ell$ only.

Previous calculations  of the background exciton-exciton
scattering~\cite{ciuti} suggest that the background interaction energy
$v_o$ is of the order of $R^*a_X^2$, where $R^*$ is the exciton Rydberg
energy 
(on the order of 10 meV for standard GaAs quantum wells) and $a_X$ is
the exciton Bohr radius (on the order of 10~nm).  As the radius $a_{B}\approx 20$nm of Wannier biexcitons is of the same order as the exciton radius $a_X$
and the Rabi energy $\Omega_R\approx 2$meV is typically of the same order as the
exciton Rydberg $R^*$, the resonant contribution $G$ is expected to
dominate over the background scattering contribution $V$ as soon as the
polariton dot is much larger than the physical size of excitons and
biexcitons $\ell \gg a_{X,B}$. As the confinement length $\ell$ of
typical polariton dots is on the order of 1~$\mu$m, this latter
condition is generally very well satisfied.

Inserting these values of the  parameters in \eq{G_dot}, one obtains a
value of $G$ on the order of $30\,\mu$eV, which is extremely promising
in view of applications: confined polariton states with a linewidth of
$70\,\mu$eV have been recently demonstrated~\cite{epfl}. On the other
hand, this value of the nonlinear coupling is safely below the energy
distance from the neighboring polariton states in the dot; the typical
energy splitting between  confined polariton states in a box potential
with a transverse size of the order of $1\mu$m are in fact on the order
of a few meV, which guarantees {\em a posteriori} the validity of our
single-mode calculations. As compared to the original
proposal~\cite{verger}, the use of the biexciton Feshbach resonance
allows us to relax the constraint on the size of the polariton dot.

\section{Few-body states}

\begin{figure*}
\begin{center}
\hspace*{0.5cm} \includegraphics[height=5cm,clip]{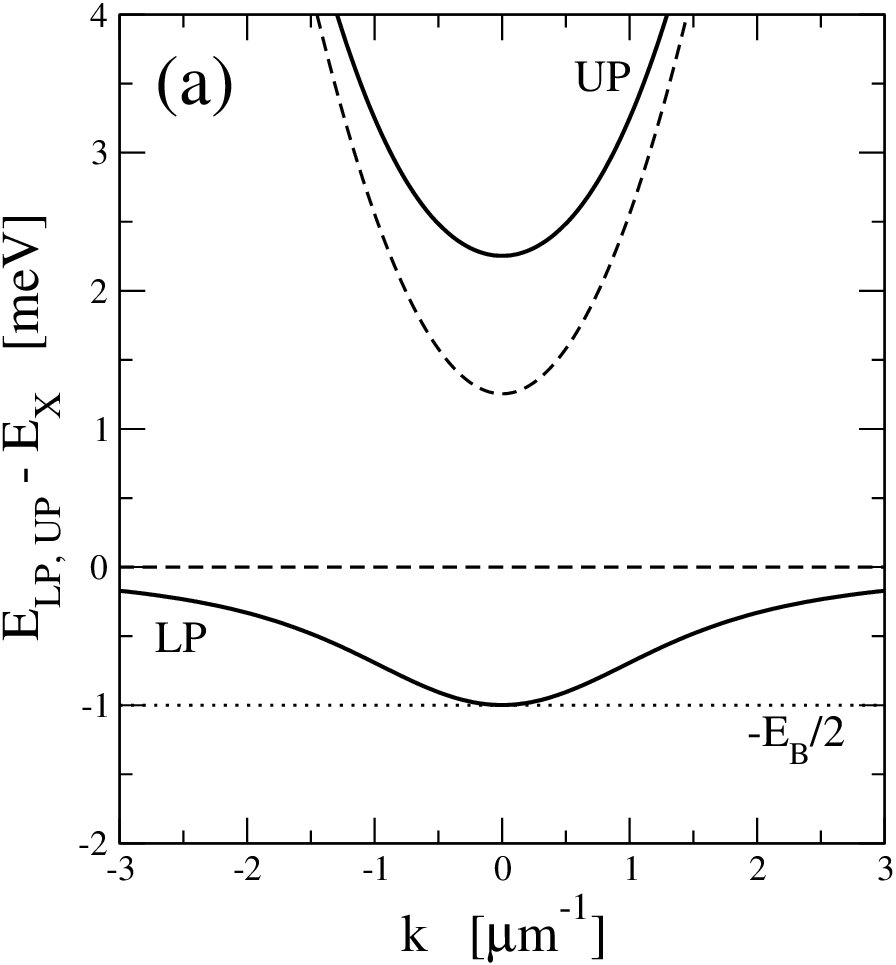}
\hspace{1.5cm} \includegraphics[height=4.5cm,clip]{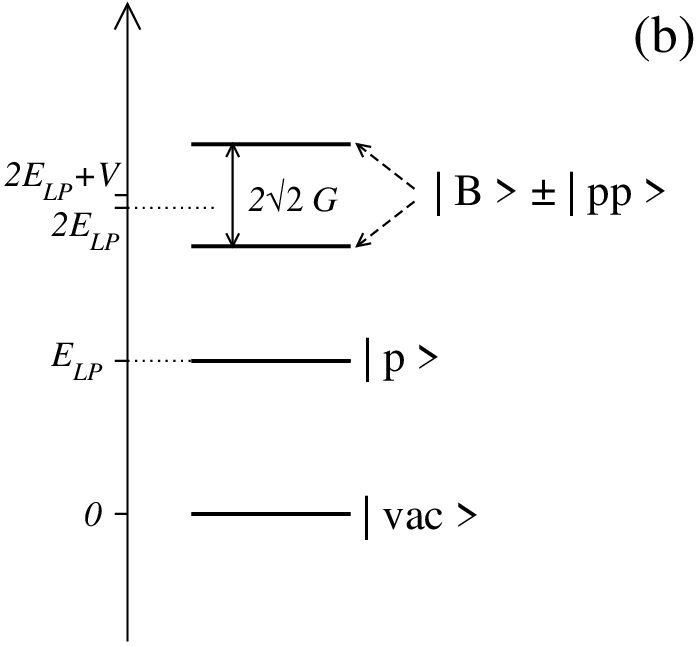} \\
\vspace{0.5cm}
\includegraphics[height=7cm,clip]{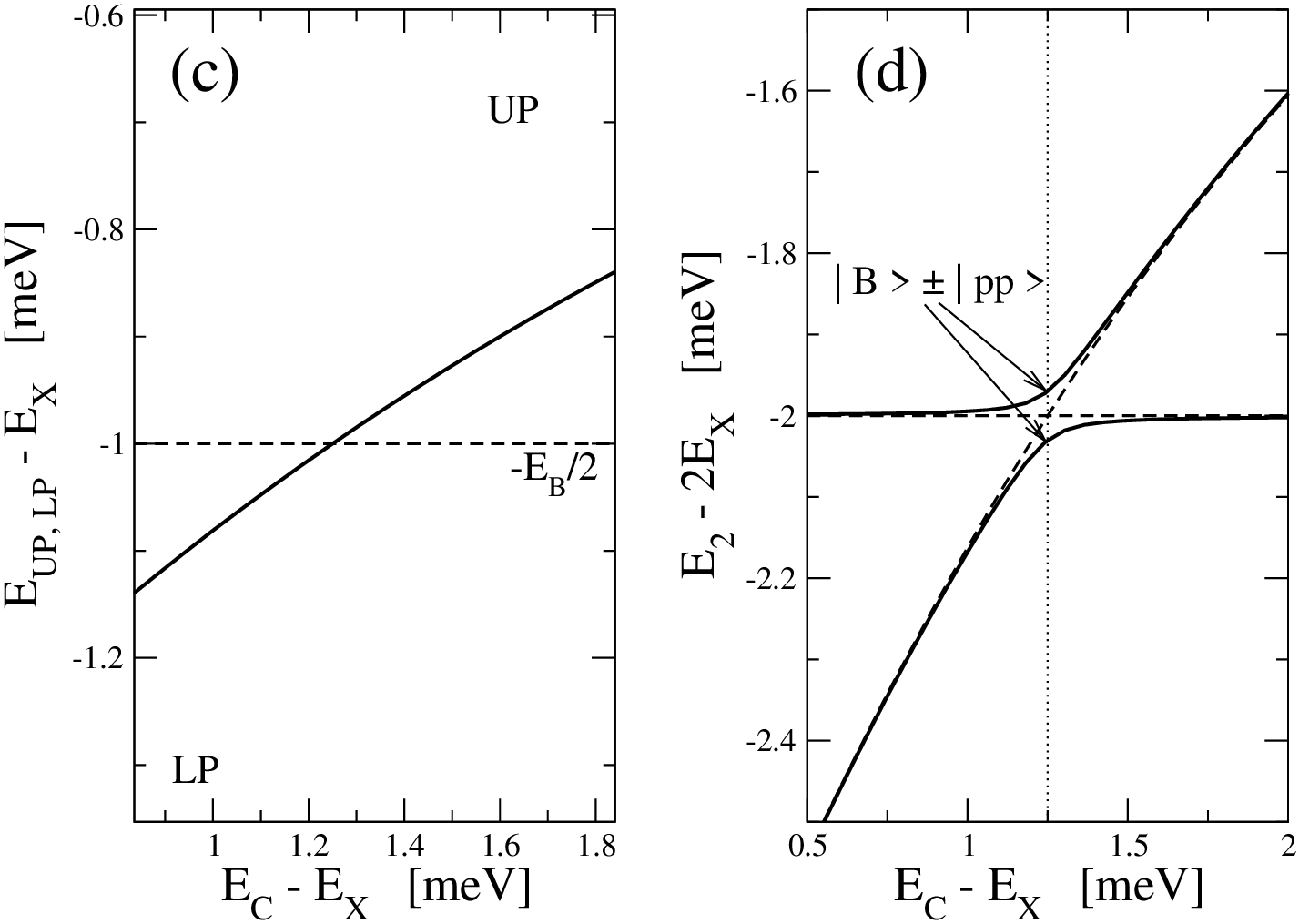}\hspace{0.5cm}
\end{center}
\caption{Upper-Left (a) panel. Solid lines: dispersion of the polariton
frequency as a function of in-plane momentum; the exciton-cavity detuning is chosen such that the lower polariton is on (Feshbach) resonance with the biexciton energy (horizontal dotted line). 
Dashed lines: bare cavity photon and exciton dispersion.  
Upper-Right (b) panel: schematic diagram of the states
belonging to the $N_{tot}=0,1,2$ manifolds exactly on Feshbach resonance
$2E_{LP}+V=E_B$.  Lower-Left (c) panel. $k=0$ anticrossing of the lower
and upper polariton branches as a function of the bare cavity frequency
$E^o_C$. The horizontal dashed line indicates half the biexciton energy;
the Feshbach resonance is located at its crossing point with the lower
polariton branch.  Lower-Right (d) panel. Position of the different
states belonging to the $N_{tot}=2$ manifold as a function of the bare
cavity frequency $E_C^o$. Dashed line: bare $|pp\rangle$ and $|B\rangle$
states. Solid lines: states originating from the mixing of the biexciton
state $|B\rangle$ and the  $|pp\rangle$  state. The
vertical dotted line indicates the Feshbach resonance point. Parameters:
$E_X^o=1.4$~meV; $\Omega_R=1.5$~meV; $G=0.03$~meV; $V\simeq 0$.  }
\label{fig:level_scheme}
\end{figure*}

As the Hamiltonian \eq{Hconcise} conserves the total number of
excitation $N_{tot}=N_{LP}+2N_{B}$, it can be easily diagonalized within
each $N_{tot}$ manifold, giving a ladder of mixed levels. An example of
this level ladder is shown in Fig.\ref{fig:level_scheme}(b) for the most
remarkable case, $2E_{LP}+V=E_B$, where the biexciton is on resonance with
the (slightly shifted) polariton states. The $N_{tot}=0$ manifold
consists of the vacuum state $|\textrm{vac} \rangle$ only, while the
$N_{tot}=1$ manifold consists of a single one-polariton state
$|p\rangle$ with energy $E_{LP}$. The $N_{tot}=2$ manifold contains of a
doublet of eigenstates which are the symmetric and antisymmetric combinations
of a biexciton $|B\rangle$ and a pair of polaritons with identical
linear polarization, $|pp\rangle$. The energies of these states are
symmetrically split by an amount $\pm\sqrt{2}G$, typically much larger than
$V$. In this way, the harmonicity of the level ladder is broken: the
transitions from the $N_{tot}=2$ to the $N_{tot}=1$ manifold occur in fact
at energies $E_{LP}\pm \sqrt{2} G$ different from the energy $E_{LP}$ of
the $N_{tot}=1$ to $N_{tot}=0$ transition.

These simple considerations straightforwardly extend to the eigenstates forming the higher $N_{tot}>2$ manifolds, which then consist of linear superpositions of states with $N_{tot}-2n$ polaritons and $n$ biexcitons ($0\le n\leq N_{tot}/2$) with suitable weights. In particular, it is interesting to note that,
differently from the atomic case, the absence of deep molecular states
other than the biexciton state should suppress all those three- and
many-body recombination processes that generally limit the lifetime of
strongly interacting atomic Bose gases~\cite{3body}.

Far from resonance $|2E_{LP}-E_B|\gg G$, the polariton and
biexciton states are no longer efficiently mixed
by the polariton-biexciton coupling and their energies approach
their uncoupled values. In this regime, the harmonicity of the level
ladder is only broken by a much weaker background scattering term of
magnitude $V\ll G$.
A complete picture of the anticrossing between the biexciton and
polariton states in the $N_{tot}=2$ manifold is shown in
Fig.\ref{fig:level_scheme}(d): while in the atomic case the difference in magnetic moment of the molecular and atomic states allows to tune their
relative energy by means of an external magnetic field, here the resonance between the polaritons and the biexciton can be obtained by suitably adjusting the cavity length. As it is shown in Fig.\ref{fig:level_scheme}(c,d), the polariton energy strongly depends on the cavity-photon energy (and in turn on the cavity length), while the biexciton energy only depends on the microscopic structure of the quantum well.

\section{Optical signatures of polariton blockade}

The pronounced anharmonicity of level scheme  of
Fig.\ref{fig:level_scheme}(b) has direct implications for the optical
properties of the polariton dot. For the sake of simplicity we restrict
here our attention to the $G \gg \Gamma, \Gamma_{add}$ limit,
where $\Gamma$ is the radiative linewidth of the polariton states and
$\Gamma_{add}$ is the additional, non-radiative contribution to the biexciton linewidth. The driving laser is assumed to be fully coherent, weak and resonant with the transition to the one polariton state $|p\rangle$, which then gets effectively populated. On the other hand, the Feshbach-resonance-induced nonlinearity shifts both states in the $N_{tot}=2$
manifold far from resonance with the laser by an amount $\sqrt{2}G$ larger than
the linewidth of the relevant states. As a consequence of this {\em
polariton blockade} effect, the $N_{tot}=2$ 
manifold remains almost unpopulated: the first polariton is able to
effectively block the entrance of the second. In a transmission geometry
where only light emitted by the polariton dot is collected, this {\em
polariton blockade} effect is clearly visible as a complete
anti-correlation between transmitted photons\footnote{If the polarization degeneracy is not lifted, the Feshbach enhancement of the polariton-polariton interactions is restricted to the singlet channel, which limits the blockade effect to polaritons of opposite polarizations. This dramatically suppresses the anti-bunching features, but leaves interesting polarization correlations in the emitted light.}.

The decay rate $\Gamma$ of the lowest energy  cavity-polaritons is
determined exclusively by the cavity-losses. While the lowest value
observed to date is $\Gamma \sim 70 \mu$eV, there are no physical
constraints that prohibit reaching much smaller values by increasing the
cavity quality factor.  On the other hand, the intrinsic biexciton decay
rate $\Gamma_{add}$ into bulk photons or surface polaritons can be
estimated to be on the order of $10 \mu$eV. Unless some specific effort
is done to suppress it in a suitably designed photonic structure, this
biexciton decay rate is most likely to set the ultimate limit to the
antibunching that is achievable in polariton dots.

A recent experiment~\cite{volz} has demonstrated  how dissipative
collisional processes offer a viable alternative route to create
strongly correlated atomic gases. The possibility of extending this idea
to the photonic case was discussed in~\cite{kiffner}. In our system, the
use of such dissipative nonlinearities provides a simple way to obtain a
strongly antibunched emission in the opposite regime $\Gamma_{add} \gg G
\ggg \Gamma$ as well: while the energy spacing between the levels
remains almost constant, the harmonicity of the level ladder in this
case is broken by the strongly varying dissipation rate. For instance,
the strong biexciton damping $\Gamma_{add}$ allows the decay rate of the
two-polariton $|pp\rangle$ state $\Gamma_2=2\Gamma+2G^2/\Gamma_{add}$ to
be much larger than the decay rate $\Gamma$ of the one-polariton
$|p\rangle$ state. As a result, the excitation of the system under a
weak and coherent pump, is limited to the $N_{tot}=1$ manifold, which
produces a stream of antibunched photons in the emission.

\section{Conclusions}
We have theoretically  investigated the physics of coupled exciton,
photon and biexciton quantum fields in planar semiconductor
microcavities. The Feshbach resonance on the biexciton state can be
exploited to dramatically enhance the amplitude of polariton scattering
processes. A strong photon antibunching is predicted for the emitted
light from strongly confined polariton dots. If our predictions are
verified experimentally, biexciton physics will become the cornerstone
for a new generation of opto-electronic devices working at the
single-photon level, with a number of applications from fundamental
science to quantum information processing.

\acknowledgments The core of this work was performed while I.C. was
visiting the  QPG group at ETH-Z\"urich. Continuous exchanges with
Michiel Wouters, Cristiano Ciuti, Giuseppe La Rocca, Andreas Reinhard,
Hakan T\"ureci, 
Antonio Badolato, Dario Gerace, and Rosario Fazio are warmly
acknowledged.

\end{document}